\documentclass[sigconf]{acmart}

\usepackage{booktabs} 
\usepackage{multirow}
\usepackage{color}
\usepackage[hyphenbreaks]{breakurl}
\usepackage{url}

\setcopyright{none}

\acmDOI{}

\acmISBN{}

\acmConference{The 17th Annual Workshop on the Economics of Information Security (WEIS)}
\acmYear{June 2018,}
\copyrightyear{\textit{Innsbruck, Austria}}

\begin{document}
\title{Ransomware Payments in the Bitcoin Ecosystem}

\author{Masarah Paquet-Clouston}
\orcid{}
\affiliation{%
  \institution{GoSecure Research}
  \city{Montreal} 
  \state{Canada} 
  \postcode{}
}
\email{mcpc@gosecure.ca}

\author{Bernhard Haslhofer}
\affiliation{%
  \institution{Austrian Institute of Technology}
  \city{Vienna} 
  \state{Austria}
  \postcode{}
}
\email{bernhard.haslhofer@ait.ac.at}

\author{Beno\^it Dupont}
\affiliation{%
  \institution{Universit\'e de Montr\'eal}
  \city{Montreal} 
  \country{Canada}
}
\email{benoit.dupont@umontreal.ca}


\begin{abstract}

Ransomware can prevent a user from accessing a device and its files until a ransom is paid to the attacker, most frequently in Bitcoin. With over 500 known ransomware families, it has become one of the dominant cybercrime threats for law enforcement, security professionals and the public. However, a more comprehensive, evidence-based picture on the global direct financial impact of ransomware attacks is still missing.
In this paper, we present a data-driven method for identifying and gathering information on Bitcoin transactions related to illicit activity based on footprints left on the public Bitcoin blockchain. We implement this method on-top-of the GraphSense open-source platform and apply it to empirically analyze transactions related to 35 ransomware families. 
We estimate the lower bound direct financial impact of each ransomware family and find that, from 2013 to mid-2017, the market for ransomware payments has a minimum worth of USD 12,768,536 (22,967.54 BTC). We also find that the market is highly skewed with only a few number of players responsible for the majority of the payments. 
Based on these research findings, policy-makers and law enforcement agencies can use the statistics provided to understand the size of the illicit market and make informed decisions on how best to address the threat.

\end{abstract}




\maketitle


\section{Introduction}\label{sec:introduction}

Ransomware attacks have eclipsed most other cybercrime threats and have become the dominant concern for law enforcement and security professionals in many nations (cf.~\cite{Europol:2017a,Scott:2016:Online,Pathak:2016a}). The device of ransomware victims are infected by a class of malicious software that, when installed on a computer, prevents a user from accessing the device --- usually through unbreakable encryption --- until a ransom is paid to the attacker. In this type of attack, cybercriminals do not profit from the resale of stolen information on underground markets to willing buyers, but from the value victims assign to their locked data and their willingness to pay a nominal fee to regain access to them. To that extent, the business model of ransomware seems conducive to more favorable monetizing opportunities than other forms of cybercrimes, due to its scalable potential and the removal of intermediaries. 

Prominent recent ransomware examples are \textit{Locky}, \textit{SamSam}, or \textit{WannaCry}, the latter infected up to 300,000 victims in 150 countries~\cite{Europol:2017a}. Like other ransomware, these families focus on extorting money from victims and thus raise fear and concern among potential victims who see the attack as a direct intimidation~\cite{Gazet:2010}. At the time of writing, there are 505\footnote{\url{https://id-ransomware.malwarehunterteam.com/}} known ransomware families detected and almost all of them demand payments in Bitcoin~\cite{Nakamoto:2008}, which is the most prominent cryptocurrency.

Yet, global and reliable statistics on the impact of cybercrime in general, and ransomware in particular, are missing, causing a large misunderstanding regarding the severity of the threat and the extent to which it fuels a large illicit business. Most of the statistics available on cybercrime and ransomware are produced by private corporations (cf.~\cite{OBrien:2017,RSA:2016:Online,Kaspersky:2016:Online}) that do not disclose their underlying methodologies and have incentives to over- or underreport them since they sell cybersecurity products and services that are supposed to protect their users against such threats~\cite{Moore:2009a}. Also, both cybercrime and ransomware attacks take place in many regions of the world and reporting the prevalence of the threat on a global level is difficult, especially when it involves a blend of fairly sophisticated technologies that may not be familiar to a large number of law enforcement organizations~\cite{Moore:2009a,Wall:2007a}. This is unfortunate because the lack of reliable statistics prevents policy-makers and practitioners from understanding the true scope of the problem, the size of the illicit market it fuels and prevents them from being able to make informed decisions on how best to address it, as well as to determine what levels of resources is needed to control it.

But ransomware offers a unique opportunity to quantify at least the direct financial impact of such threat: ransomware payments are transferred in Bitcoin, which is a peer-to-peer cryptocurrency with a public transaction ledger --- known as \textit{blockchain} --- that is shared among peers. When ransomware payments can be identified correctly, the Bitcoin blockchain provides a reliable basis on which to assess ransomware cash flows. Furthermore, a number of clustering heuristics (cf.~\cite{Reid:2013a,Meiklejohn:2013a,Monaco:2015a}) have been proposed that support partitioning the set of Bitcoin addresses observed in the entire cryptocurrency ecosystem into maximal subsets, which are likely controlled by the same real-world actor. Previous studies have measured ransomware payments in the ecosystem, but focused on a single ransomware family (CryptoLocker~\cite{Liao:2016a}), did not make use of known clustering heuristics~\cite{Kharraz:2015a} or, at the time of this writing, disclosed limited information on their underlying methodology~\cite{Bursztein:2017a}.

In order to provide a more comprehensive picture of the global direct financial impact of ransomware attacks, we propose a data-driven method for identifying and gathering information on Bitcoin transactions related to ransomware and then apply this method for 35 ransomware families. More specifically, the contributions of this paper can be summarized as follows:

\begin{itemize}

	\item We propose a data-driven method for identifying and gathering Bitcoin transactions, related to ransomware attacks, that goes beyond known clustering heuristics.
    
    \item We implement this method on-top-of the open-source GraphSense cryptocurrency analytics platform\footnote{\url{http://graphsense.info/}} and make the transaction extraction\footnote{\url{https://github.com/behas/ransomware-dataset}} and analytics procedures\footnote{\url{https://github.com/behas/ransomware-analytics}} openly available.
	
    \item We apply the method on a sample of 35 different ransomware families and find new addresses related to each ransomware family, distinguish collectors from payment addresses and, when possible, track where the money is cashed out.
    
    \item We quantify the lower direct financial impact of each ransomware family, show how ransom payments evolve over time and find that from 2013 to mid-2017, the market for ransomware payments for 35 families sums to a minimum amount of USD 12,768,536 (22,967.54 BTC).
    
\end{itemize}

To our knowledge, this paper is the first to present a method to assess payments of a large number of ransomware families in Bitcoin and to provide a lower bound for their direct financial impacts, while being openly available and reproducible. Our proposed method and findings also roughly correspond with concurrent research reported in Bursztein et al. \cite{Bursztein:2017a} and Huang et al.~\cite{Huang:2018aa}.


The remainder of this paper is organized as follows: we provide further details on ransomware and traceability of Bitcoin transactions in Section~\ref{sec:related_work}. Our methodology for identifying and gathering Bitcoin transactions is described in Section~\ref{sec:methodology} and the results of our study is presented in Section~\ref{sec:results}. The discussion follows in Section~\ref{sec:discussion} along with the conclusion in Section~\ref{sec:conclusions}.

\section{State of the Art}\label{sec:related_work}

\subsection{Ransomware}

The concept of extorting money from user devices through malicious means has had a long existence, such as fake anti-virus that forced users to buy a software to erase an inexistent malware from their  devices~\cite{Pathak:2016a,Hampton:2015a,Song:2016a}. Still, ransomware is a criminal innovation that seeks to monetize illegally accessed information by charging its rightful owner a ransom --- usually a few hundred dollars --- to recover the personal files that have a unique sentimental or administrative value. 

Nowadays, two modes of attacks have been used by ransomware authors to prevent file access on a device. The first mode of attack aims at locking out a user from a device by disabling the operating system (OS). When the user starts the device, a ransom note appears requesting money to be transferred for the device to start as usual~\cite{Song:2016a,Hampton:2015a}. The second mode of attack is more sophisticated and uses cryptography. The technique is to encrypt a user's files on a device before requesting a ransom in exchange for the key that will decrypt them~\cite{Song:2016a}. 

Since the first implementation of encryption as an attack technique, other technologies have been leveraged to increase the efficiency of new variants of ransomware. The Onion Routing (Tor) Protocol has allowed ransomware attackers to use an anonymous and direct communication channel with their victims. The use of cryptocurrencies for ransom payments has enabled relatively anonymous money exchanges, while evading the control of established financial institutions and their law enforcement partners. The combination of strong and well-implemented cryptographic techniques to take files hostage, the Tor protocol to communicate anonymously, and the use of a cryptocurrency to receive unmediated payments provide altogether a high level of impunity for ransomware attackers ~\cite{Owen:2015}.

Many argue that ransomware authors have proved to be highly innovative in the past years. Since 2013 and the first introduction of the Cryptolocker ransomware, new variants  have been designed and distributed by ambitious cybercriminals, building on the success of previous versions or fixing previous errors ~\cite{Hampton:2015a,Kharraz:2015a,Pathak:2016a}. Yet, focusing on the speed at which ransomware authors modify their malware and the technologies used may lead to overestimate the severity of the threat. 

As the current hype would have it, ransomware authors would make large amounts of money --- up to millions of dollars --- with this successful online black mailing activity ~\cite{OBrien:2017,CyberThreatAlliance:2015:Online,Hernandez:2017a}. As it is often the case, the reality is not that simple. In 2015, Kharraz et al. ~\cite{Kharraz:2015a} published a long-term study on ransomware attacks in which they analyzed 1,359 samples from 15 ransomware families. Even though ransomware has evolved, these authors found that \emph{the number of families with sophisticated destructive capabilities remains quite small}. They also found that malware authors mostly used superficial techniques to encrypt or delete a victim's files.  Flaws were, moreover, found in the code, making the attack easily defeated. Similarly, Gazet ~\cite{Gazet:2010} conducted a comparative analysis of 15 ransomware and discovered that the code used was often basic and built on high-level languages. Looking at the victims and the ransoms asked, the author concluded that ransomware attackers followed a low-cost/low-risk business model: they did not aim at mass extortion, but relied instead on small attacks for small ransoms, which could be compensated by mass propagation. 

Moreover, although ransomware was perceived, at first, as a destructive form of attack almost impossible to prevent and detect, many initiatives led by the security community have tempered this initial assessment ~\cite{Kharraz:2015a}. For example, Kharraz et al. ~\cite{Kharraz:2016a}, Scaife et al. ~\cite{Scaife:2016a}, Song et al. ~\cite{Song:2016a}, Continella et al. ~\cite{Continella2016a} and Kolodenker et al. ~\cite{kolodenker:2017a} all developed tools to detect ransomware-like behaviors and prevent them from successfully encrypting a device. These tools help mitigate ransomware attacks, minimizing the potential damages caused by this threat. The ransomware threat is thus certainly evolving and growing, but is not out of control. The community keeps finding ways to detect and block it preemptively. Moreover, when a user is infected, an international initiative called "No More Ransom!"\footnote{\url{https://www.nomoreransom.org/}} provides decryption tools for victims of ransomware. These tools were developed by exploiting technical flaws in malware implementations and, at the moment, more than 40 of them are available on the website for different ransomware strains.

\subsection{Bitcoin Traceability Research}

Bitcoin is a peer-to-peer cryptocurrency initially introduced by Satoshi Nakamoto (a pseudonym) in 2008~\cite{Nakamoto:2008}. It can be used to execute pseudo-anonymous payments globally within a short period of time and --- at least before the enormous rise in popularity at the end of 2017 --- with comparably low transaction costs. All executed and confirmed financial transactions are stored in a shared and transparent ledger, known as the blockchain, which is publicly accessible. Each transaction is represented by a list of inputs pointing back to outputs and a list of outputs, each reflecting an amount of Bitcoins transferred to a specific recipient's address. A Bitcoin address is an alphanumeric string derived from the public key of an asymmetric key pair generated by a Bitcoin user. Every user can hold multiple key pairs (and addresses) in a wallet, and is encouraged to use a new address for each transaction to increase the level of anonymity.

A number of heuristics have been developed to analyze transactions and group all addresses in the Bitcoin blockchain into maximal subsets (clusters) that can be associated with some real-world actors. The \textit{multiple-input heuristics}~\cite{Nakamoto:2008,Reid:2013a} takes into account that two addresses used as inputs in the same transaction must be controlled by the same real-world actor. If one input address is used in another transaction along with other input addresses, they can all be linked to the same real-world actor. Cluster identification can further be refined by applying \textit{change heuristics}~\cite{Meiklejohn:2013a,Spagnuolo:2014b,Androulaki:2013e}, which exploit the concept of "change addresses" in Bitcoin. If one pays 1.5 BTC for a service, but has an address with 2 BTC, the remaining 0.5 BTC will be sent back to the user using what is defined as a "change address". 

When clusters are correlated with attribution data (tags) from external sources, such as publicly available information in forums (cf.~\cite{Fleder:2015a}) or specific sites (e.g., \url{blockchain.info}, \url{walletexplorer.org}), it is possible to deanonymize large fractions of the entire Bitcoin transaction network. Clustering of Bitcoin addresses and tagging addresses with attribution data are two central features that are nowadays supported by modern cryptocurrency analytics tools (e.g., Chainalysis, Elliptic, GraphSense). 

Applying these strategies on public transactions turns Bitcoin into --- at most --- a pseudo-anonymous currency, in which monetary flows can be traced from one known or unknown address to another. These strategies can identify Bitcoin addresses and clusters related to illicit activities, unless one makes use of \emph{mixing} or \emph{CoinJoin} services. Mixing Services --- also known as \emph{tumblers} --- are specialized intermediaries that break the link between senders and receivers by mixing coins and transactions with those of other users (c.f.~\cite{Moser:2013aa}). A CoinJoin transaction, on the other hand, is a special transaction in which multiple senders and recipients of funds combine their payments in a single aggregated transaction. This requires a dedicated service (e.g., JoinMarket) that matches interested users and supports them in creating the transaction~\cite{Moser:2016aa}. Both types of services facilitate the amalgamation of coins belonging to multiple individuals in a single transaction, making the tracing of illicit activity more difficult.

The effectiveness of clustering heuristics has been investigated by Nick~\cite{Nick:2015a}, who assessed the well-known multiple-input clustering heuristics on a ground-truth dataset of approximately 37K wallets and found that such a clustering algorithm can guess, on average, 68.59\% of all addresses belonging to a wallet. Building on that, Harrigan and Fretter~\cite{Harrigan:2016a} concluded that address clustering in the Bitcoin network was effective due to identified address reuse and the existence of superclusters with incremental growth (e.g., exchanges, gambling sites, darknet marketplaces).

\subsection{Tracing Bitcoin Transactions related to Ransomware}

A ransomware attacker who requests payments in Bitcoin will broadcast a Bitcoin address to which the victim needs to send money to. This address is a ransom payment address from which clustering heuristics in the Bitcoin network can be computed. Three previous studies have investigated ransomware activity in the Bitcoin network. Kharraz et al.~\cite{Kharraz:2015a} analyzed 1,872 Bitcoin addresses related to the CryptoLocker ransomware. They concluded that Bitcoin addresses related to Cryptolocker had similar transaction records, such as a short activity period and a few numbers of small transactions. In total, 84\% of the addresses analyzed had no more than six transactions and 69\% were active for less than 10 days. Liao et al.~\cite{Liao:2016a} also performed a measurement analysis of the Cryptolocker ransomware. They started their investigation with two Bitcoin addresses and generated a cluster of 968 addresses. They filtered transactions based on ransom amounts and time and provided a lower and upper bound for Cryptolocker's economy. They mentioned that possible connections exist between this ransomware and Bitcoin services, such as Bitcoin Fog and BTC-e, and other cybercrime activities, like darknet markets. Finally, a concurrent research reported in Bursztein et al. ~\cite{Bursztein:2017a} and  Huang et al. ~\cite{Huang:2018aa} traced Bitcoin transactions of several ransomware families.  The research estimated that about USD 16 million ransomware payments were made with Bitcoins over a two-year period. 

This study goes beyond the state of the art on ransomware and Bitcoin traceability research by presenting a simple automated method, built on known clustering heuristics, to systematically trace monetary flows. It applies the method on Bitcoin transactions related to 35 ransomware families to identify, quantify and compare their financial activity in the Bitcoin network.


\section{Methodology}\label{sec:methodology}
 
In the following section, we describe how we identify, collect, and filter payments related to ransomware attacks by analyzing the Bitcoin blockchain.

\subsection{Seed Dataset Collection}\label{subsec:datset_collection}

To begin, Bitcoin addresses related to ransomware attacks were collected from various sources. A total of 7,037 addresses related to the Locky ransomware were provided to us by the Anti-Phishing Working Group (APWG)\footnote{\url{https://www.antiphishing.org/}}. An additional 139 Bitcoin addresses were found in a thread maintained by Michael Gillepsi\footnote{Michael Gillepsi is the creator of the initiative : \url{https://id-ransomware.malwarehunterteam.com/}}. Through additional online searches, 46 Bitcoin addresses were found in various sources, such as security researchers' blogs or websites of organizations analyzing ransomware activity. In total, we extracted 7,222 Bitcoin addresses related to 67 ransomware families. Throughout the whole study, we refer to them as \textit{seed addresses} because they are the ones used to generate the larger dataset. 

\subsection{Bitcoin Network Construction}\label{subsec:network_construction}

We extracted transaction data from the Bitcoin blockchain using the GraphSense open-source platform. Our most recent expansion ran on October 28\textsuperscript{th}, 2017 with 489,181 blocks, 260,167,622 transactions and 312,506,384 addresses.

In order to trace monetary flows, we computed two types of network representations over the entire blockchain: the \textit{address graph}, in which each vertex represents a Bitcoin address and each directed edge represents the aggregated set of transactions transferring value from one address to another. For each directed edge we computed summary statistics, such as the number of transactions and the estimated value flow between two addresses, considering the daily Bitcoin/USD closing price as conversion rates. The technical details of these computations are described in more details in an earlier paper~\cite{Haslhofer:2016a}. Moreover, transaction outputs containing explicit change addresses (i.e., addresses that were also referenced by one of the inputs within the same transaction), were removed to eliminate monetary flows having the same address as source and destination.

The second type of network representation is the \textit{cluster graph}. To compute this graph, we partitioned the set of addresses observed in the entire blockchain into maximal subsets (\textit{clusters}) that are likely to be controlled by the same real-world actor using the well-known~\cite{Reid:2013a} and efficient~\cite{Harrigan:2016a} \textit{multiple-input clustering heuristics}. The underlying intuition is that if two addresses (i.e.: A and B) are used as inputs in the same transaction while one of these addresses along with another address (i.e: B and C) are used as inputs in another transaction, then the three addresses (A, B and C) must somehow be controlled by the same  real-world actor~\cite{Meiklejohn:2013a}, who conducted both transactions and therefore possesses the private keys corresponding to all three addresses. In the \textit{cluster graph}, the nodes represent address clusters and the directed edges represent transactions between clusters. Since each cluster represents an aggregation of addresses, the edges between clusters can be seen as an aggregation of each transaction value taking into account USD conversion rates.

In order to associate real-world actors, such as Bitcoin exchanges or gambling sites, with addresses and clusters, we gather publicly available information, so-called \emph{tags}, from two main external sources: \url{walletexplorer.com} and \url{blockchain.info}. Each tag associates a specific Bitcoin address with some contextually relevant information (e.g., \textit{BTC-e.com}) about real-world actors and facilitates the interpretation of monetary flows. The great power of Bitcoin address attribution lies in its combination with clustering heuristics: if one can attribute a single address within a cluster containing hundred of thousands of addresses, one can attribute the entire cluster. When investigating monetary flows, Bitcoin exchanges are of great interest because they are the entry and exit points of the cryptocurrency ecosystem where fiat currencies (e.g., USD, EUR) are converted into cryptocurrencies and vice versa.

\subsection{Dataset Expansion Procedure }\label{subsec:datset_expansion}

To expand the seed address dataset, which was obtained as described in Section \ref{subsec:datset_collection}, we matched the set of seed addresses with the set of all addresses extracted from the blockchain. This eliminated 100 seed addresses not appearing in the blockchain because they have not (yet) received ransom payments from victims and have therefore not been used in a Bitcoin transaction, reducing our dataset to 7,122 addresses from 38 families. We then expanded the dataset by linking these seed addresses to their corresponding clusters in the cluster graph, which was pre-computed through the multiple-input heuristics. We refer to these addresses as \textit{expanded addresses}.

However, if a ransomware author was involved in other activities that implied Bitcoin transactions before the ransomware campaign, the multiple-input heuristic could result in false positives. Thus, to ensure that the addresses in the expanded dataset were related to ransomware activity, we applied a time filter on the expanded dataset by determining a start date of ransomware campaigns. For 25 families, we used the Google trend searches and extracted the first month in which online searches about the ransomware family took place. Google trend searches can be a good indicator of the beginning of a ransomware campaign because individuals or organizations hit by a ransomware campaign are likely to search online to learn more about the threat before they decide on a course of action. This method was, however, not successful for 13 ransomware families from which Google trend search did not have any data. For those cases, we looked for online articles or blogs on the ransomware family and took the earliest article published on the subject, no matter in which language it was written. Out of the 13 families, we did not find any information on the start date of three of them because no articles or blogs were published related to them; they were sometimes only listed as a potential threat among other ransomware families.

Our final sample contains 7,118 addresses related to 35 ransomware families and corresponding time filters (see Table~\ref{tab:time_filters}). In the remainder of this paper, due to limited space, the subsequent tables will display the Top 15 ransomware families\footnote{The results for the 35 families can be reproduced with the scripts and the datasets provided  in the Github repositories}. 
\begin{table}
\centering
\resizebox{\columnwidth}{!}{%
\begin{tabular}{rlll}
  \toprule
 	& Family
    & Ransomware Start Date
    & Investigation Method \\ 
  \hline
  1 & Locky & 2016-02 & Google Trends \\ 
  2 & CryptXXX & 2016-04 & Google Trends \\ 
  3 & CryptoLocker & 2013-09 & Google Trends \\ 
  4 & DMALockerv3 & 2016-01 & Google Trends \\ 
  5 & CryptoTorLocker2015 & 2015-02 & Google Trends \\ 
  6 & Globe & 2013-04 & Google Trends \\ 
  7 & SamSam & 2016-01 & Google Trends \\ 
  8 & NoobCrypt & 2015-12 & Manual search \\ 
  9 & EDA2 & 2015-09 & Manual search \\ 
  10 & Flyper & 2016-09 & Manual search \\ 
  11 & Globev3 & 2017-01 & Manual search \\ 
  12 & JigSaw & 2016-04 & Google Trends \\ 
  13 & Cryptohitman & 2016-05 & Google Trends \\ 
  14 & TowerWeb & 2016-06 & Manual search \\ 
  15 & WannaCry & 2017-05 & Google Trends \\ 
   \hline
\end{tabular}
}
\caption{Time filters applied for top 15 ransomware families.}
\label{tab:time_filters}
\end{table}
Table~\ref{tab:dataset_summary} summarizes the top 15 ransomware families ordered by the number of addresses in our expanded dataset after application of time filters (\textit{Exp. Addr. (TF)}). It also lists the number of collected seed addresses (\textit{Seed Addr.}), the number of expanded addresses before time filtering (\textit{Exp. Addr.}), and the number of clusters (\textit{Clusters}) that can be assigned to each ransomware family. The numbers in Table~\ref{tab:dataset_summary} show that the multiple-input heuristics can identify a large number of Bitcoin addresses related to ransomware attacks. Table~\ref{tab:dataset_summary} also shows that the seed address distribution is highly skewed.


In the case of \textit{Locky}, we found that the number of seed addresses is almost equal to the number of expanded addresses because the multiple-input heuristic was already computed on the seed addresses provided by the APWG. Also, the number of \textit{CryptoLocker} addresses corresponds exactly to the number of addresses (968) reported by Liao et al.~\cite{Liao:2016a} in an earlier study. We take these observations as a validation of our expansion method and its implementation.
\begin{table}
\centering
\resizebox{\columnwidth}{!}{%
\begin{tabular}{rlllll}
  \toprule
 	& Family
    & Seed Addr.
    & Clusters
    & Exp. Addr.
    & Exp. Addr. (TF) \\ 
  \midrule
  1 & Locky & 7,038 & 1 & 7,094 & 7,093 \\ 
  2 & CryptXXX & 1 & 1 & 1,742 & 1,742 \\ 
  3 & CryptoLocker & 2 & 1 & 968 & 968 \\ 
  4 & DMALockerv3 & 9 & 3 & 165 & 165 \\ 
  5 & CryptoTorLocker2015 & 1 & 1 & 159 & 121 \\ 
  6 & Globe & 8 & 2 & 87 & 87 \\ 
  7 & SamSam & 44 & 11 & 47 & 47 \\ 
  8 & NoobCrypt & 2 & 1 & 28 & 28 \\ 
  9 & EDA2 & 2 & 2 & 33 & 26 \\ 
  10 & Flyper & 2 & 1 & 26 & 26 \\ 
  11 & Globev3 & 9 & 3 & 19 & 18 \\ 
  12 & JigSaw & 12 & 4 & 17 & 17 \\ 
  13 & Cryptohitman & 1 & 1 & 14 & 13 \\ 
  14 & TowerWeb & 1 & 1 & 14 & 8 \\ 
  15 & WannaCry & 5 & 1 & 6 & 6 \\ 
  \bottomrule
\end{tabular}
}
\caption{Dataset statistics for top 15 ransomware families.}
\label{tab:dataset_summary}
\end{table}
When looking at Table~\ref{tab:dataset_summary}, one can observe that time filtering does not eliminate many addresses in the expanded dataset. This indicates that the multiple-input clustering already delivers addresses within the expected time frame of each ransomware campaign.

\subsection{Beyond the Clustering Process: Tracing Outgoing Relationships }

The dataset expansion using the multiple-input heuristic points to new addresses related to ransomware attacks. While investigating the expanded dataset, we developed a simple method to go beyond the clustering process and trace monetary flows. Indeed, by focusing on outgoing transactions for one ransomware family, one can find common addresses receiving money from the expanded addresses related to that ransomware family. 

\begin{figure}
	\includegraphics[width=\columnwidth]{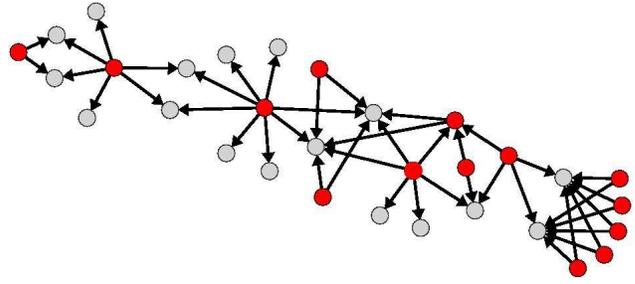}
	\caption{CryptoHitman Outgoing-Relationships Graph}
    \label{fig:crytohitman}
\end{figure}

The method consists of taking into account, for each expanded address, all the outgoing transactions and their respective outputs. With this, an \textit{outgoing-relationships} graph can be built for each ransomware family. The nodes in the graph are either the expanded addresses or the addresses receiving money from the expanded addresses. The edges in the graph illustrate the direction of the monetary flow. For example, Figure~\ref{fig:crytohitman} illustrates the \textit{outgoing-relationship} graph for the \textit{CryptoHitman} ransomware family. The red nodes represent addresses from our expanded address dataset which belong to the \textit{CryptoHitman} family and the gray nodes represent output addresses not in the dataset. The graph shows that some addresses are key since they receive, more than once, money from known \textit{CryptoHitman} addresses. Other gray addresses in the graph only receive one incoming transaction. They could possibly be related to the \textit{CryptoHitman} ransomware, but the information in the graph is insufficient to allow such conclusion. 
While the \textit{CryptoHitman} graph is small enough for visual inspections, other ransomware families have large graphs and require automated mechanisms to distinguish  key addresses. Thus, to automatically distinguish key addresses in an {outgoing-relationships} graph, we develop a simple method. For each node, we calculate the number of incoming relationships in the graph, as expressed in Definition~\ref{def:indegree}:

\begin{definition}\label{def:indegree}

The in-degree $deg^{-}$ of a Bitcoin address $a$ is the sum of all unique incoming relationships of $a$ within the scope of a family-specific \textit{outgoing-relationships} graph.

\end{definition}

We consider that each node that has $deg(a)^{-} \geq 2$, in a family-specific \textit{outgoing-relationships} graph, is a \emph{key address} for this ransomware family. Even though some addresses are already in the dataset while others are not, they are all identified as key addresses related to the ransomware family.

We computed an \textit{outgoing-relationships} graph for each family in the dataset and calculated the metric by applying the above definition. We found, in total, 2,077 key addresses from the 35 families studied. Table~\ref{tab:key_addresses} presents the number of key addresses found with the \textit{outgoing-relationships} graph for each family, and shows how many were already part of our expanded dataset (\textit{Key Expanded Addr.}) and how many were added by this method (\textit{New Key Addr.}). 
\begin{table}
\centering
\resizebox{\columnwidth}{!}{%
\begin{tabular}{rlll}
   \toprule
 	& Family
    & New Key Addr. 
    & Key Expanded Addr.\\ 
  \midrule
  1 & CryptXXX & 488 & 438 \\ 
  2 & Locky & 305 & 266\\ 
  3 & CryptoTorLocker2015 & 160 & 37  \\ 
  4 & DMALockerv3 & 53 & 18 \\ 
  5 & Globe & 47 &38 \\ 
  6 & NoobCrypt & 43 & 11 \\ 
  7 & SamSam & 31 & 6\\
  8 & CryptoLocker & 26 & 24 \\ 
  9 & EDA2 & 16 & 3 \\ 
  10 & JigSaw & 16 & 1 \\ 
  11 & Cryptohitman & 9 & 1\\ 
  12 & TowerWeb & 9 & 1 \\ 
  13 & Globev3 &6 & 0 \\ 
  14 & Flyper & 5 & 3 \\ 
  15 & VenusLocker & 5 & 1 \\
 \bottomrule
\end{tabular}
}
\caption{Key Addresses identified for each family.}
\label{tab:key_addresses}
\end{table}

\subsection{Estimating the Lower Bound Financial Impact of Each Ransomware Family}
With the dataset generated through the different steps mentioned above, an assessment of the minimum direct financial impact of each ransomware family is possible. The multiple-input clustering heuristic allowed an expansion of the dataset and the time filtering ensured that the expanded addresses were within the time frame of each ransomware campaign. Also, the method of tracing outgoing relationships found key addresses that received money from the expanded addresses related to a ransomware family. The key addresses already in the expanded dataset (red nodes) are filtered out of the expanded dataset for the financial assessment, in order to avoid double-counting ransom payments.

\section{The Impact of Ransomware}\label{sec:results}

Building on the methodology presented in the previous section and the resulting dataset, we can now analyze Bitcoin transactions related to ransomware. In the following section, we report our findings on tracing ransomware monetary flows. Then, we provide a lower bound estimation for the direct financial impact of the Top 15 families in our dataset and give insight into the value and longitudinal development of ransomware payments. Lastly, we present an estimation of the minimum worth of the market for ransomware payments. 

\subsection{Following the Money Trace}

By computing the \textit{outgoing-relationships} graph for each ransomware family and applying the condition mentioned above, key addresses for each ransomware family were found. Although the minimum for an address to be determined as key was to score $deg(a)^{-} \geq 2$, many key addresses had a much higher score. Within the sample of 2,077 key addresses, the average $deg(a)^{-}$ was 12 (std=27.66) incoming relationships and the median was 6. The maximum $deg(a)^{-}$ in the sample went up to 742 incoming relationships. This indicates that ransomware authors do tend to consolidate their money into one or several key addresses.

\begin{figure}
	\includegraphics[width=\columnwidth]{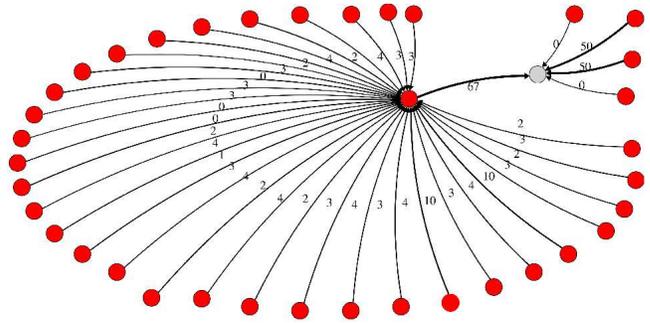}
	\caption{Locky collector address example.}
	\label{fig:locky_collector}
\end{figure}

Intuitively, these key addresses can be considered \textit{collectors} of a ransomware family. We define a collector as an address used to collect or aggregate payments from several payment addresses. To picture the role of a collector, Figure~\ref{fig:locky_collector} shows the relationships of a subset of \textit{Locky} addresses. It illustrates that an address that was already in the expanded dataset (red node) has a high degree centrality and receives 32 payments of less or equal to 10 BTC. Considering the high degree centrality, this address can be considered a collector of the \textit{Locky} ransomware family. That figure also shows that the high-degree centrality address sends 67 Bitcoins to a gray address, which is an address not in the expanded dataset. Similarly, two other addresses, from the expanded dataset, send 50 Bitcoins to that gray address. At a higher level, this gray address can also be considered a collector of the \textit{Locky} ransomware family. 

However, it must be noted that a collector address does not necessarily belong to the same cluster of a family's seed and expanded addresses (such as the gray node in Figure~\ref{fig:locky_collector}). This is because deciding whether an address is a collector or not depends on the monetary flow in an \textit{outgoing-relationships} graph related to a ransomware family and has nothing to do with the multiple-input heuristic results, which is based on the author having the private keys of all addresses in the cluster. Indeed, some collector addresses can rather be part of a larger cluster representing Bitcoin exchange services or gambling sites, which can be used by attackers to convert ransom payments to fiat currencies or to camouflage monetary flows.

If a key address belongs to a large known cluster, it could then be considered the end route of tracing ransomware payments. As explained before in the methodology, such assessment is possible by investigating tags associated with addresses and address clusters. We investigated the tags associated to the 2,077 key addresses and their corresponding clusters in more detail and found 163 key addresses related to 28 tagged clusters with additional contextual information. Of these 163 collectors, 86 were related to known exchanges organizations, such as \textit{BTC-e.com}, \textit{LocalBitcoin.com}, \textit{Kraken.com} and \textit{Xapo.com}. Another 47 were related to gambling sites like \textit{SatoshiDice.com}, \textit{Bitzillions.com}, \textit{SatoshiMines.com}, \textit{BetCoin Dice} and \textit{FortuneJack.com}. A total of 12 addresses were linked to mixing services, such as \textit{BitcoinFog.info} and \textit{Helix Mixer}. These services are specialized intermediaries which mix coins and transactions of different actors and thereby camouflage the digital trace of cryptocurrency transactions. They play a central role in money laundering and cybercrime-related activities that rely on cryptocurrencies as a payment method.

Although our information on real-world actors behind addresses and clusters was limited to the tags we retrieved from external sources and therefore incomplete, we found that some ransomware attackers directly sent the ransom payments to known actors, mostly gambling and exchange services. We also found that some ransomware families specifically transacted multiple times with the same actor. For example, 20 \textit{CryptoTorLocker2015} key addresses were related to the \textit{SatoshiDice} organization and 25 \textit{Locky} key addresses were linked to the \textit{BTC-e} exchange. Also, about 27 key addresses from five ransomware families belonged to the \textit{Localbitcoin.com}\footnote{https://localbitcoins.com/about} cluster, which is an exchange that allows individuals to buy and sell Bitcoins to people who are geographically close.  

As extra information, the \textit{outgoing-relationships} analysis also linked some families together. It illustrated that the \textit{Globe} and \textit{Globev3} families sent money to the same untagged collector address, which was to be expected based on their shared naming features, but was confirmed through our methodology. Similarly, 10 key addresses, with a few number of transactions and no tags, received money from both the \textit{TowerWeb} and \textit{Cryptohitman} addresses. Intuitively, we can assume that these two families might be related to the same real-world actors who may run two families of ransomware simultaneously or may launder money on behalf of the two different groups.

\subsection{Lower Bound Direct Financial Impacts}\label{subsec:revenues}

Besides tracing ransomware monetary flows, we assessed the lower bound financial impact of each ransomware family. The basis for our estimation was the time-filtered expanded ransomware dataset described in Section~\ref{subsec:datset_expansion}. In order to avoid double-counting of ransomware payments, we removed known collector addresses from the dataset. Table~\ref{tab:Received_payments} presents the total amount of received payments for the Top 15 ransomware families in the dataset. It shows received payments in Bitcoin (BTC), rounded to two decimal places, and in U.S dollars (USD). 
\begin{table}[t]
\centering
\begin{tabular}{rllll}
  \hline
 & Family & Addresses & BTC & USD \\ 
  \hline
  1 & Locky & 6,827 & 15,399.01 & 7,834,737 \\ 
  2 & CryptXXX & 1,304 & 3,339.68 & 1,878,696 \\ 
  3 & DMALockerv3 & 147 & 1,505.78 & 1,500,630 \\ 
  4 & SamSam & 41 & 632.01 & 599,687 \\ 
  5 & CryptoLocker & 944 & 1,511.71 & 519,991 \\ 
  6 & GlobeImposter & 1 & 96.94 & 116,014 \\ 
  7 & WannaCry & 6 & 55.34 & 102,703 \\ 
  8 & CryptoTorLocker2015 & 94 & 246.32 & 67,221 \\ 
  9 & APT & 2 & 36.07 & 31,971 \\ 
  10 & NoobCrypt & 17 & 54.34 & 25,080 \\ 
  11 & Globe & 49 & 33.03 & 24,319 \\ 
  12 & Globev3 & 18 & 14.34 & 16,008 \\ 
  13 & EDA2 & 23 & 7.1 & 15,111 \\ 
  14 & NotPetya & 1 & 4.39 & 11,458 \\ 
  15 & Razy & 1 & 10.75 & 8,073 \\ 
   \hline
\end{tabular}
\caption{Received payments per ransom family (Top 15).}
\label{tab:Received_payments}
\end{table}
We find that the ransomware family that generated the largest direct financial impact in our dataset is \textit{Locky}, which received payments totalizing USD 7,834,737. The second ransomware family is \textit{CryptXXX} with a lower bound direct financial impact of USD 1,878,696, followed by the \textit{DMALockerv3} ransomware family with USD 1,500,630. Based on our dataset, these are the three families that created a lower bound direct financial impact of more than one million. Then, \textit{SamSam}, \textit{Cryptolocker} and \textit{GlobeImposter} generated lower bound direct financial impacts of hundreds of thousands of dollars each. As we go down the ranking, a rapid decline is observed: the ransomware occupying the 15th position, \textit{Razy}, barely gathered a lower bound of USD 8,073. 

Due to the worth of the Bitcoin being highly volatile, we do not consider these amounts as representing ransomware revenue. Indeed, such assumption would assume that ransomware authors cashed out immediately after receiving victims' payments, which may not be the case. 

Also, when comparing the amounts above with findings reported in other studies, we observe similarities and discrepancies. The results for \textit{Locky} and \textit{CryptXXX} are consistent with the concurrent research reported in Huang et al.~\cite{Huang:2018aa} and  Bursztein et al.~\cite{Bursztein:2017a}. These authors found that the \textit{Locky} ransomware generated a direct financial impact of approximately USD 7,8 million and the \textit{CryptXXX} ransomware approximately USD 1.9 million. However, there is a discrepancy in the results for \textit{CryptoLocker}: they estimated that \textit{Cryptolocker} created roughly USD 2 million in direct financial impact versus USD 519,991 in our study.  Liao et al.~\cite{Liao:2016a} measured \textit{CryptoLocker} payments from September 2013 until January 2014 and reported a lower bound direct financial impact of USD 310,472 and an upper bound of USD 539,080, which is much closer to our result. Yet, the discrepancy seems to come from the addition of a single additional seed address --- disclosed in the Huang et al.~\cite{Huang:2018aa} study --- that led to an expanded cluster of 3,489 addresses. This cluster neither appears in our research nor in Liao et al.'s. 

Another discrepancy lies in the result, displayed in  Bursztein et al.~\cite{Bursztein:2017a}, about the \textit{SamSam} ransomware: USD 1.9 million for this research against USD 583,498 in this study. The differences may arise from the different number of seed addresses used in the Bursztein et al. research. Finally, we identify high or moderate performing ransomware families, such as \textit{DMALockerv3} and \textit{NoobCrypt}, that did not register in the concurrent research.

\subsection{Inspecting Payments}\label{subsec:payments}

\begin{figure}
	\includegraphics[width=\columnwidth]{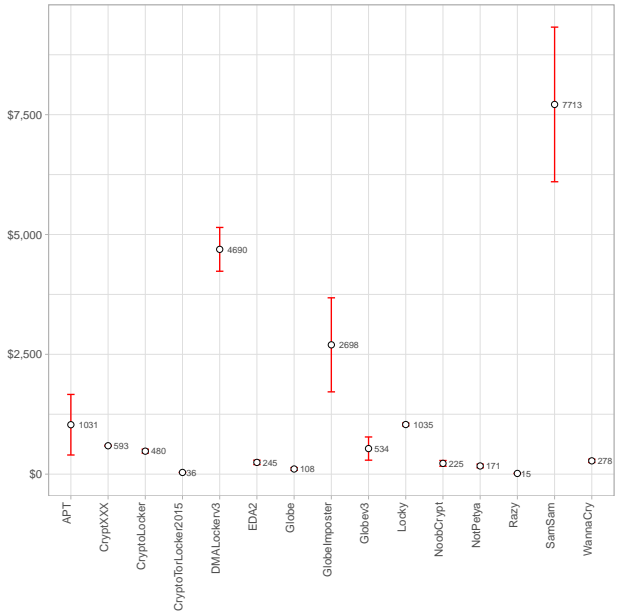}
	\caption{Mean payment per family with standard mean errors.}
    \label{fig:mean_payments}
\end{figure}

Figure~\ref{fig:mean_payments} presents the mean payment per family (and the standard mean errors) of the Top 15 families. It shows that the incoming transactions of 12 ransomware families range from very low payments up to USD 2,000. Three ransomware families have higher payments on average: \textit{DMALockerv3}, \textit{GlobeImposter} and \textit{SamSam}. In January 2016, \textit{DMALockerv3} was known to ask for ransom payments of 15 BTC (which was equivalent to \$6491.25)~\cite{OBrien:2017,Meskauskas:2017:Online}. The \textit{SamSam} ransomware was also known to ask ransoms based on the number of machines infected and the ransom could go from 1.7 BTC (\$4,600) to decrypt a given machine up to 12 BTC (\$32,800) to decrypt all machines infected ~\cite{Doman:2017:Online}. For the \textit{GlobeImposter} ransomware, however, we could not find a justification for the relative high mean payment value and mean error rate. We only identified a single address for that ransomware family in our dataset and, therefore, could not compute means across addresses belonging to that family.
\begin{figure}
	\includegraphics[width=\columnwidth]{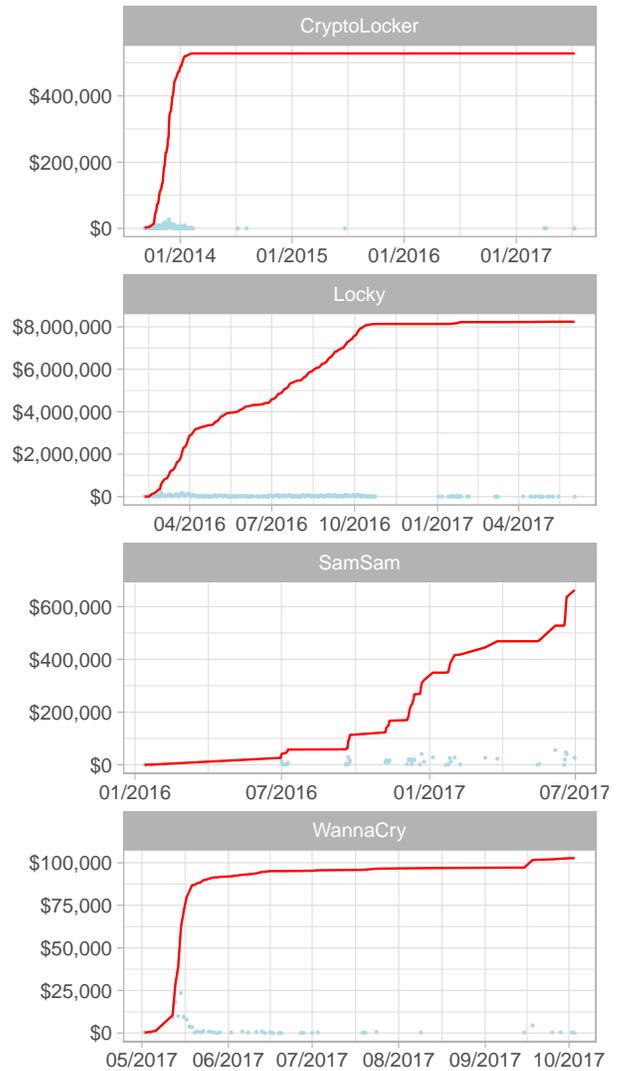}
	\caption{Longitudinal payment trend per family.}
    \label{fig:longitudinal_trends}
\end{figure}

Figure~\ref{fig:longitudinal_trends} shows cumulative (red line) and non-cumulative ransom payments (blue dots) over time for a selection of four ransomware families: \textit{Cryptolocker}, \textit{Locky}, \textit{SamSam} and \textit{Wannacry}. For three famous families, \textit{CryptoLocker}, \textit{Locky} and \textit{Wannacry}, it shows the viral effect of ransomware attacks and ransom payments. It also illustrates that famous ransomware campaigns are likely to be a short-term, one-time deal, in which a ransomware author makes money quickly and then stops, possibly due to various forms of security interventions. However, the \textit{SamSam} ransomware seems to behave differently since the cumulative payment curve shows a somewhat linear trend over a whole year, from July 2016 to July 2017. The difference in this campaign could be related to the different approach used by the ransomware authors, which is known to be more targeted ~\cite{Doman:2017:Online}.

\subsection{Market for Ransomware Payments}\label{sec:overall_market}
When summing the lower bound direct financial impacts of all 35 families analyzed in our study, we find that, from 2013 to mid-2017, the minimum worth of the market for ransom payments represents USD 12,768,536 (22,967.54 BTC). This means that the \textit{Locky} ransomware accounts for more than 50\% of the ransomware payments and the first three families account for 86\% of the market while the other 32 families share the remaining 12\%. These results are similar to the concurrent research reported in Bursztein et al. ~\cite{Bursztein:2017a} and Huang et al.~\cite{Huang:2018aa}, which also conclude that the ransomware market is dominated by a few kingpins. 


\section{Discussion}\label{sec:discussion}

Overall, we believe that the method presented in this paper led to novel insights for each ransomware family. Ransom payment addresses and collectors were differentiated in the dataset, allowing one to assess ransomware lower bound direct financial impacts without double-counting. Plus, we were able to trace monetary flows of ransomware payments and identify destinations, such as Bitcoin exchanges or gambling services, when contextually related information (tags) was available. Our method is reproducible and could be repeated for additional families with an updated seed dataset. Plus, computation of address clusters over the most recent state of the Bitcoin blockchain, along with more identification of clusters belonging to specific groups, could greatly increase the knowledge on exit points of ransomware monetary flows.

We are well aware that our approach has a number of limitations. First, our methodology relies on a set of seed addresses manually collected and the effectiveness of the multiple-input heuristics for uncovering previously unknown addresses linked to this family. Thus, it misses other ransomware families as well as other addresses that might belong to the same family, but cannot be linked to the same cluster. Still, the more addresses from various families become available, the more accurate the picture of the overall market for ransom payments will become. We address this limitation by constraining our analysis to ``lower bound'' direct financial impacts, to ensure we are not claiming to assess the total impacts of a ransomware family or of the entire market for ransom payments.

Second, our approach is limited by the extent and quality of the attribution data (tags) available. Without this information, clusters remain anonymous and inferences about their real-world nature are impossible. Nevertheless, we believe that such data will increasingly become available in the near future with the growing popularity of cryptocurrencies and analytics tools. 

Third, tumblers or mixing services, which facilitate the amalgamation of coins belonging to multiple individuals in a single transaction, increase the difficulty of tracing monetary flows in the Bitcoin network (cf.,~\cite{Moser:2013a}). We believe that our methodology is robust to such services because it only considers payments to addresses derived from a manually collected set of ransomware payment addresses and their direct outgoing neighbors in the address graph. Thus, in the worst case, a key address would represent the entry point of a mixing service. 

We also note that the transactions we attribute to ransomware families could be part of CoinJoin transactions . However, we argue that matching transactions with those of other users when collecting ransom payments would add an undesirable third party (the CoinJoin service) dependency in the process. This should hardly be implemented in practice as using CoinJoin services  to collect ransoms would also create delays in payments and certainly cause considerable technical efforts for ransomware attackers. This assumption is somehow confirmed by Huang et al.~\cite{Huang:2018aa}, who applied known CoinJoin detection heuristics on their dataset and did not find such transactions. 

Despite these limitations, we have shown that one can uncover valuable insights into ransomware payments and the market values of these attacks. Through the analysis of 35 ransomware families in the Bitcoin network, we find that there are some clear inequalities in the market, which could be considered as a top-heavy market in which only a few players are responsible for most of the ransom payments. This is also in line with the concurrent research reported in 
Huang et al. ~\cite{Huang:2018aa} and Burzstein et al. ~\cite{Bursztein:2017a}. Such finding has implications for law enforcement agencies seeking to disrupt this market: mobilizing their limited resources on a small number of highly capable players could lead to takedowns and have a major (negative) impact on the ransomware economy.

Moreover, when masking major ransomware families, such as \textit{Locky}, \textit{CryptXXX} and \textit{DMALockerv3}, the drop in ransom amounts is substantial and we find that more than half of the ransomware family in the sample is responsible for less than USD 8,000 of direct financial impacts. 
Kharraz et al.~\cite{Kharraz:2015a} who studied 1,359 samples from 15 ransomware families and Gazet~\cite{Gazet:2010}, who reversed-engineered 15 ransomware samples, both found that most ransomware families used superficial and flawed techniques to encrypt files. Few of them had actual destructive capabilities and most of them could be easily defeated. This could explain why only few ransomware families succeed at generating ransom payments worth millions.

Such observations do not mean that the ransomware threat should be underestimated. Although the minimum worth of the market for ransom payments, taking into account 35 families, is a relatively modest amount (about USD 12 million) compared to the hype surrounding the issue, the overall direct and indirect damages they caused to individual and organizational victims are much higher ~\cite{CyberThreatAlliance:2015:Online}. Yet, there is not doubt that initiatives developed by the community to prevent ransomware attacks ~\cite{Kharraz:2016a,Scaife:2016a,Song:2016a}, as well as the initiative "No More Ransom!"\footnote{\url{https://www.nomoreransom.org/}}, that make ransomware decryption tools available to victims, can have a positive impact on limiting ransom payments. Some of the ransomware families in our datasets have decryption tools available on this community website. Although this could explain why some families do not have a large direct financial impact, further analysis should look into the performance changes of a ransomware family once a decryption tool is made available.


\section{Conclusions}\label{sec:conclusions}


We present a novel method for identifying and gathering information on Bitcoin transactions related to illicit activity. We implement this method on-top-of the GraphSense open-source platform and apply it to empirically analyze transactions related to 35 ransomware families. We estimate the lower bound direct financial impact of each ransomware family and find that, from 2013 to mid-2017, the  market for ransomware payments has a minimum worth of USD 12,768,536 (22,967.54 BTC). We also find that the market is highly skewed, dominated by a few number of players. From these findings, we conclude that the total ransom amounts gathered through ransomware attacks are relatively low compared to the hype surrounding this issue. 

We believe that our simple data-driven methodology and findings provide valuable insights and carry implications for security companies, government agencies and the public in general. It could, for instance, be adopted in threat intelligence systems for following ransomware payments associated with new campaigns in real time, and for identifying inflection points such as explosive growth phases and slowdown periods, when the plateau of ransom payments is reached.

An evidence-based and more granular longitudinal tracking of the entire ransomware economy would allow government agencies and security companies to fine-tune their intervention efforts and awareness campaigns to focus on the two or three most active and dynamic threats. In other words, by making more reliable, comprehensive, and timely information available on the nature and scope of the ransomware problem, our methodology can help lead the discussion on how best to address the threat at scale and support subsequent decision-making.

One straightforward future work would be to extend our analysis to additional ransomware families. Work in that direction should also take into account the emergence of post-Bitcoin cryptocurrencies, such as Monero, Ethereum or Zcash, that have advanced privacy features and are gaining popularity in the digital underground~\cite{Europol:2017a}. \textit{Kirk} is the first ransomware family that has been reported to use Monero for ransom payments ~\cite{Abrams:2017:Online}. 

Another possible area of future work lies in the application of this methodology on other illicit activities that channel their financial transactions through the Bitcoin network, such as other extortion cases, trafficking of illicit goods or money laundering. Since Bitcoin is nowadays "accounting for over 40\% of all identified criminal-to-criminal payments" and cryptocurrencies seem to "[...] establish themselves as single common currency for cybercriminals[...]"~\cite{Europol:2015a}, there are plenty of application areas for such a method.

\section*{Acknowledgments}

This work was partly funded by the European Commission through the project TITANIUM (Project ID: 740558). We thank Dave Jevans and Mike D'Ambrogia from the APWG organization for contributing to the seed addresses dataset.


\bibliographystyle{ACM-Reference-Format}
\bibliography{bibliography}


\begin{thebibliography}{39}


\ifx \showCODEN    \undefined \def \showCODEN     #1{\unskip}     \fi
\ifx \showDOI      \undefined \def \showDOI       #1{#1}\fi
\ifx \showISBNx    \undefined \def \showISBNx     #1{\unskip}     \fi
\ifx \showISBNxiii \undefined \def \showISBNxiii  #1{\unskip}     \fi
\ifx \showISSN     \undefined \def \showISSN      #1{\unskip}     \fi
\ifx \showLCCN     \undefined \def \showLCCN      #1{\unskip}     \fi
\ifx \shownote     \undefined \def \shownote      #1{#1}          \fi
\ifx \showarticletitle \undefined \def \showarticletitle #1{#1}   \fi
\ifx \showURL      \undefined \def \showURL       {\relax}        \fi
\providecommand\bibfield[2]{#2}
\providecommand\bibinfo[2]{#2}
\providecommand\natexlab[1]{#1}
\providecommand\showeprint[2][]{arXiv:#2}

\bibitem[\protect\citeauthoryear{Abrams}{Abrams}{2016}]%
        {Abrams:2017:Online}
\bibfield{author}{\bibinfo{person}{Lawrence Abrams}.}
  \bibinfo{year}{2016}\natexlab{}.
\newblock \bibinfo{title}{Star Trek Themed Kirk Ransomware Brings us Monero and
  a Spock Decryptor!}
\newblock   (\bibinfo{date}{March} \bibinfo{year}{2016}).
\newblock
\showURL{%
\url{https://www.bleepingcomputer.com/news/security/star-trek-themed-kirk-ransomware-brings-us-monero-and-a-spock-decryptor/}}


\bibitem[\protect\citeauthoryear{Alliance}{Alliance}{2015}]%
        {CyberThreatAlliance:2015:Online}
\bibfield{author}{\bibinfo{person}{Cyber~Threat Alliance}.}
  \bibinfo{year}{2015}\natexlab{}.
\newblock \bibinfo{booktitle}{{\em Lucrative Ransomware Attacks: Analysis of
  the CryptoWall Version 3 Threat}}.
\newblock \bibinfo{type}{{T}echnical {R}eport}. \bibinfo{institution}{Cyber
  Threat Alliance}.
\newblock
\showURL{%
\url{http://cdn2.hubspot.net/hubfs/343670/images/resources/cryptowall-report.pdf?t=1463166454836}}


\bibitem[\protect\citeauthoryear{Androulaki, Karame, Roeschlin, Scherer, and
  Capkun}{Androulaki et~al\mbox{.}}{2013}]%
        {Androulaki:2013e}
\bibfield{author}{\bibinfo{person}{Elli Androulaki}, \bibinfo{person}{Ghassan~O
  Karame}, \bibinfo{person}{Marc Roeschlin}, \bibinfo{person}{Tobias Scherer},
  {and} \bibinfo{person}{Srdjan Capkun}.} \bibinfo{year}{2013}\natexlab{}.
\newblock \showarticletitle{Evaluating User Privacy in Bitcoin}. In
  \bibinfo{booktitle}{{\em International Conference on Financial Cryptography
  and Data Security}}. Springer, \bibinfo{pages}{34--51}.
\newblock


\bibitem[\protect\citeauthoryear{Bursztein, McRoberts, and
  Invernizzi}{Bursztein et~al\mbox{.}}{2017}]%
        {Bursztein:2017a}
\bibfield{author}{\bibinfo{person}{Elie Bursztein}, \bibinfo{person}{Kylie
  McRoberts}, {and} \bibinfo{person}{Luca Invernizzi}.}
  \bibinfo{year}{2017}\natexlab{}.
\newblock \showarticletitle{Tracking Desktop Ransomware Payments}.
  \bibinfo{publisher}{Presented at Black Hat USA, Las Vegas, United States}.
\newblock


\bibitem[\protect\citeauthoryear{Continella, Guagnelli, Zingaro, De~Pasquale,
  Barenghi, Zanero, and Maggi}{Continella et~al\mbox{.}}{2016}]%
        {Continella2016a}
\bibfield{author}{\bibinfo{person}{Andrea Continella},
  \bibinfo{person}{Alessandro Guagnelli}, \bibinfo{person}{Giovanni Zingaro},
  \bibinfo{person}{Giulio De~Pasquale}, \bibinfo{person}{Alessandro Barenghi},
  \bibinfo{person}{Stefano Zanero}, {and} \bibinfo{person}{Federico Maggi}.}
  \bibinfo{year}{2016}\natexlab{}.
\newblock \showarticletitle{Shieldfs: a self-healing, ransomware-aware
  filesystem}. In \bibinfo{booktitle}{{\em Proceedings of the 32nd Annual
  Conference on Computer Security Applications}}. ACM,
  \bibinfo{pages}{336--347}.
\newblock


\bibitem[\protect\citeauthoryear{Doman}{Doman}{2017}]%
        {Doman:2017:Online}
\bibfield{author}{\bibinfo{person}{Chris Doman}.}
  \bibinfo{year}{2017}\natexlab{}.
\newblock \bibinfo{title}{SamSam Ransomware Targeted Attacks Continue}.
\newblock   (\bibinfo{date}{June} \bibinfo{year}{2017}).
\newblock
\showURL{%
\url{https://www.alienvault.com/blogs/labs-research/samsam-ransomware-targeted-attacks-continue}}


\bibitem[\protect\citeauthoryear{Europol}{Europol}{2015}]%
        {Europol:2015a}
\bibfield{author}{\bibinfo{person}{Europol}.} \bibinfo{year}{2015}\natexlab{}.
\newblock \bibinfo{booktitle}{{\em Internet Organised Crime Threat Assessment
  (IOTA)}}.
\newblock \bibinfo{type}{{T}echnical {R}eport}. \bibinfo{institution}{Europol}.
\newblock


\bibitem[\protect\citeauthoryear{Europol}{Europol}{2017}]%
        {Europol:2017a}
\bibfield{author}{\bibinfo{person}{Europol}.} \bibinfo{year}{2017}\natexlab{}.
\newblock \bibinfo{booktitle}{{\em Internet Organised Crime Threat Assessment
  (IOTA)}}.
\newblock \bibinfo{type}{{T}echnical {R}eport}. \bibinfo{institution}{Europol}.
\newblock


\bibitem[\protect\citeauthoryear{Fleder, Kester, and Pillai}{Fleder
  et~al\mbox{.}}{2015}]%
        {Fleder:2015a}
\bibfield{author}{\bibinfo{person}{Michael Fleder}, \bibinfo{person}{Michael~S
  Kester}, {and} \bibinfo{person}{Sudeep Pillai}.}
  \bibinfo{year}{2015}\natexlab{}.
\newblock \showarticletitle{Bitcoin Transaction Graph Analysis}.
\newblock \bibinfo{journal}{{\em arXiv preprint arXiv:1502.01657\/}}
  (\bibinfo{year}{2015}).
\newblock


\bibitem[\protect\citeauthoryear{Gazet}{Gazet}{2010}]%
        {Gazet:2010}
\bibfield{author}{\bibinfo{person}{Alexandre Gazet}.}
  \bibinfo{year}{2010}\natexlab{}.
\newblock \showarticletitle{Comparative Analysis of Various Ransomware Virii}.
\newblock \bibinfo{journal}{{\em Journal in computer virology\/}}
  \bibinfo{volume}{6}, \bibinfo{number}{1} (\bibinfo{year}{2010}),
  \bibinfo{pages}{77--90}.
\newblock


\bibitem[\protect\citeauthoryear{Hampton and Baig}{Hampton and Baig}{2015}]%
        {Hampton:2015a}
\bibfield{author}{\bibinfo{person}{Nikolai Hampton} {and}
  \bibinfo{person}{Zubair~A Baig}.} \bibinfo{year}{2015}\natexlab{}.
\newblock \showarticletitle{Ransomware: Emergence of the Cyber-Extortion
  Menace}.
\newblock  (\bibinfo{year}{2015}).
\newblock


\bibitem[\protect\citeauthoryear{Harrigan and Fretter}{Harrigan and
  Fretter}{2016}]%
        {Harrigan:2016a}
\bibfield{author}{\bibinfo{person}{Martin Harrigan} {and}
  \bibinfo{person}{Christoph Fretter}.} \bibinfo{year}{2016}\natexlab{}.
\newblock \showarticletitle{The Unreasonable Effectiveness of Address
  Clustering}. In \bibinfo{booktitle}{{\em Ubiquitous Intelligence \&
  Computing, Advanced and Trusted Computing, Scalable Computing and
  Communications, Cloud and Big Data Computing, Internet of People, and Smart
  World Congress, 2016 Intl IEEE Conferences}}. IEEE,
  \bibinfo{pages}{368--373}.
\newblock


\bibitem[\protect\citeauthoryear{Haslhofer, Karl, and Filtz}{Haslhofer
  et~al\mbox{.}}{2016}]%
        {Haslhofer:2016a}
\bibfield{author}{\bibinfo{person}{Bernhard Haslhofer}, \bibinfo{person}{Roman
  Karl}, {and} \bibinfo{person}{Erwin Filtz}.} \bibinfo{year}{2016}\natexlab{}.
\newblock \showarticletitle{O Bitcoin Where Art Thou? Insight into Large-Scale
  Transaction Graphs}. In \bibinfo{booktitle}{{\em SEMANTiCS (Posters,
  Demos)}}.
\newblock


\bibitem[\protect\citeauthoryear{Hernandez-Castro, Cartwright, and
  Stepanova}{Hernandez-Castro et~al\mbox{.}}{2017}]%
        {Hernandez:2017a}
\bibfield{author}{\bibinfo{person}{Julio Hernandez-Castro},
  \bibinfo{person}{Edward Cartwright}, {and} \bibinfo{person}{Anna Stepanova}.}
  \bibinfo{year}{2017}\natexlab{}.
\newblock \showarticletitle{Economic Analysis of Ransomware}.
\newblock  (\bibinfo{year}{2017}).
\newblock


\bibitem[\protect\citeauthoryear{Huang, McCoy, Aliapoulios, Li, Invernizzi,
  Bursztein, McRoberts, Levin, Levchenko, and Snoeren}{Huang
  et~al\mbox{.}}{2018}]%
        {Huang:2018aa}
\bibfield{author}{\bibinfo{person}{Danny~Yuxing Huang}, \bibinfo{person}{Damon
  McCoy}, \bibinfo{person}{Maxwell~Matthaios Aliapoulios},
  \bibinfo{person}{Vector~Guo Li}, \bibinfo{person}{Luca Invernizzi},
  \bibinfo{person}{Elie Bursztein}, \bibinfo{person}{Kylie McRoberts},
  \bibinfo{person}{Jonathan Levin}, \bibinfo{person}{Kirill Levchenko}, {and}
  \bibinfo{person}{Alex~C Snoeren}.} \bibinfo{year}{2018}\natexlab{}.
\newblock \showarticletitle{Tracking Ransomware End-to-end}. In
  \bibinfo{booktitle}{{\em Tracking Ransomware End-to-end}}. IEEE,
  \bibinfo{pages}{0}.
\newblock


\bibitem[\protect\citeauthoryear{Kharraz, Arshad, Mulliner, Robertson, and
  Kirda}{Kharraz et~al\mbox{.}}{2016}]%
        {Kharraz:2016a}
\bibfield{author}{\bibinfo{person}{Amin Kharraz}, \bibinfo{person}{Sajjad
  Arshad}, \bibinfo{person}{Collin Mulliner}, \bibinfo{person}{William~K
  Robertson}, {and} \bibinfo{person}{Engin Kirda}.}
  \bibinfo{year}{2016}\natexlab{}.
\newblock \showarticletitle{UNVEIL: A Large-Scale, Automated Approach to
  Detecting Ransomware.}. In \bibinfo{booktitle}{{\em USENIX Security
  Symposium}}. \bibinfo{pages}{757--772}.
\newblock


\bibitem[\protect\citeauthoryear{Kharraz, Robertson, Balzarotti, Bilge, and
  Kirda}{Kharraz et~al\mbox{.}}{2015}]%
        {Kharraz:2015a}
\bibfield{author}{\bibinfo{person}{Amin Kharraz}, \bibinfo{person}{William
  Robertson}, \bibinfo{person}{Davide Balzarotti}, \bibinfo{person}{Leyla
  Bilge}, {and} \bibinfo{person}{Engin Kirda}.}
  \bibinfo{year}{2015}\natexlab{}.
\newblock \showarticletitle{Cutting the Gordian Knot: A Look under the Hood of
  Ransomware Attacks}. In \bibinfo{booktitle}{{\em International Conference on
  Detection of Intrusions and Malware, and Vulnerability Assessment}}.
  Springer, \bibinfo{pages}{3--24}.
\newblock


\bibitem[\protect\citeauthoryear{Kolodenker, Koch, Stringhini, and
  Egele}{Kolodenker et~al\mbox{.}}{2017}]%
        {kolodenker:2017a}
\bibfield{author}{\bibinfo{person}{Eugene Kolodenker}, \bibinfo{person}{William
  Koch}, \bibinfo{person}{Gianluca Stringhini}, {and} \bibinfo{person}{Manuel
  Egele}.} \bibinfo{year}{2017}\natexlab{}.
\newblock \showarticletitle{PayBreak: Defense against cryptographic
  ransomware}. In \bibinfo{booktitle}{{\em Proceedings of the 2017 ACM on Asia
  Conference on Computer and Communications Security}}. ACM,
  \bibinfo{pages}{599--611}.
\newblock


\bibitem[\protect\citeauthoryear{Liao, Zhao, Doup{\'e}, and Ahn}{Liao
  et~al\mbox{.}}{2016}]%
        {Liao:2016a}
\bibfield{author}{\bibinfo{person}{Kevin Liao}, \bibinfo{person}{Ziming Zhao},
  \bibinfo{person}{Adam Doup{\'e}}, {and} \bibinfo{person}{Gail-Joon Ahn}.}
  \bibinfo{year}{2016}\natexlab{}.
\newblock \showarticletitle{Behind Closed Doors: Measurement and Analysis of
  CryptoLocker Ransoms in Bitcoin}. In \bibinfo{booktitle}{{\em Electronic
  Crime Research (eCrime), 2016 APWG Symposium on}}. IEEE,
  \bibinfo{pages}{1--13}.
\newblock


\bibitem[\protect\citeauthoryear{Meiklejohn, Pomarole, Jordan, Levchenko,
  McCoy, Voelker, and Savage}{Meiklejohn et~al\mbox{.}}{2013}]%
        {Meiklejohn:2013a}
\bibfield{author}{\bibinfo{person}{Sarah Meiklejohn}, \bibinfo{person}{Marjori
  Pomarole}, \bibinfo{person}{Grant Jordan}, \bibinfo{person}{Kirill
  Levchenko}, \bibinfo{person}{Damon McCoy}, \bibinfo{person}{Geoffrey~M
  Voelker}, {and} \bibinfo{person}{Stefan Savage}.}
  \bibinfo{year}{2013}\natexlab{}.
\newblock \showarticletitle{A Fistful of Bitcoins: Characterizing Payments
  among Men with no Names}. In \bibinfo{booktitle}{{\em Proceedings of the 2013
  conference on Internet measurement conference}}. ACM,
  \bibinfo{pages}{127--140}.
\newblock


\bibitem[\protect\citeauthoryear{Meskauskas}{Meskauskas}{2017}]%
        {Meskauskas:2017:Online}
\bibfield{author}{\bibinfo{person}{Thomas Meskauskas}.}
  \bibinfo{year}{2017}\natexlab{}.
\newblock \bibinfo{title}{DMA Locker Ransomware}.
\newblock   (\bibinfo{date}{May} \bibinfo{year}{2017}).
\newblock
\showURL{%
\url{https://www.pcrisk.com/removal-guides/9702-dma-locker-ransomware}}


\bibitem[\protect\citeauthoryear{Monaco}{Monaco}{2015}]%
        {Monaco:2015a}
\bibfield{author}{\bibinfo{person}{John~V Monaco}.}
  \bibinfo{year}{2015}\natexlab{}.
\newblock \showarticletitle{Identifying Bitcoin Users by Transaction Behavior}.
  In \bibinfo{booktitle}{{\em SPIE Defense+ Security}}. International Society
  for Optics and Photonics, \bibinfo{pages}{945704--945704}.
\newblock


\bibitem[\protect\citeauthoryear{Moore, Clayton, and Anderson}{Moore
  et~al\mbox{.}}{2009}]%
        {Moore:2009a}
\bibfield{author}{\bibinfo{person}{Tyler Moore}, \bibinfo{person}{Richard
  Clayton}, {and} \bibinfo{person}{Ross Anderson}.}
  \bibinfo{year}{2009}\natexlab{}.
\newblock \showarticletitle{The Economics of Online Crime}.
\newblock \bibinfo{journal}{{\em The Journal of Economic Perspectives\/}}
  \bibinfo{volume}{23}, \bibinfo{number}{3} (\bibinfo{year}{2009}),
  \bibinfo{pages}{3--20}.
\newblock


\bibitem[\protect\citeauthoryear{M{\"o}ser}{M{\"o}ser}{2013}]%
        {Moser:2013a}
\bibfield{author}{\bibinfo{person}{Malte M{\"o}ser}.}
  \bibinfo{year}{2013}\natexlab{}.
\newblock \showarticletitle{Anonymity of Bitcoin Transactions}. In
  \bibinfo{booktitle}{{\em M{\"u}nster bitcoin conference}}.
  \bibinfo{pages}{17--18}.
\newblock


\bibitem[\protect\citeauthoryear{M{\"o}ser and B{\"o}hme}{M{\"o}ser and
  B{\"o}hme}{2016}]%
        {Moser:2016aa}
\bibfield{author}{\bibinfo{person}{Malte M{\"o}ser} {and}
  \bibinfo{person}{Rainer B{\"o}hme}.} \bibinfo{year}{2016}\natexlab{}.
\newblock \showarticletitle{Join me on a market for anonymity}. In
  \bibinfo{booktitle}{{\em Workshop on Privacy in the Electronic Society}}.
\newblock


\bibitem[\protect\citeauthoryear{M{\"o}ser, B{\"o}hme, and Breuker}{M{\"o}ser
  et~al\mbox{.}}{2013}]%
        {Moser:2013aa}
\bibfield{author}{\bibinfo{person}{M. M{\"o}ser}, \bibinfo{person}{R.
  B{\"o}hme}, {and} \bibinfo{person}{D. Breuker}.}
  \bibinfo{year}{2013}\natexlab{}.
\newblock \showarticletitle{An inquiry into money laundering tools in the
  Bitcoin ecosystem}. In \bibinfo{booktitle}{{\em 2013 APWG eCrime Researchers
  Summit}}. \bibinfo{pages}{1--14}.
\newblock
\showDOI{%
\url{https://doi.org/10.1109/eCRS.2013.6805780}}


\bibitem[\protect\citeauthoryear{Nakamoto}{Nakamoto}{2008}]%
        {Nakamoto:2008}
\bibfield{author}{\bibinfo{person}{Satoshi Nakamoto}.}
  \bibinfo{year}{2008}\natexlab{}.
\newblock \bibinfo{title}{Bitcoin: A peer-to-peer electronic cash system}.
\newblock   (\bibinfo{year}{2008}).
\newblock
\newblock
\shownote{Available at: \url{https://bitcoin.org/bitcoin.pdf}.}


\bibitem[\protect\citeauthoryear{Nick}{Nick}{2015}]%
        {Nick:2015a}
\bibfield{author}{\bibinfo{person}{Jonas~David Nick}.}
  \bibinfo{year}{2015}\natexlab{}.
\newblock {\em \bibinfo{title}{Data-Driven De-Anonymization in Bitcoin}}.
\newblock \bibinfo{thesistype}{Master's\ thesis}. \bibinfo{school}{ETH Zurich}.
\newblock
\showDOI{%
\url{https://doi.org/10.3929/ethz-a-010541254}}
\newblock
\shownote{Master's Thesis, Distributed Computing Group Computer Engineering and
  Networks Laboratory, ETH Z{\"u}rich, August 9, 2015.}


\bibitem[\protect\citeauthoryear{O'Brien}{O'Brien}{2017}]%
        {OBrien:2017}
\bibfield{author}{\bibinfo{person}{Dick O'Brien}.}
  \bibinfo{year}{2017}\natexlab{}.
\newblock \bibinfo{title}{Ransomware 2017, An ISTR Special Report}.
\newblock
  \bibinfo{howpublished}{\url{https://www.symantec.com/content/dam/symantec/docs/security-center/white-papers/istr-ransomware-2017-en.pdf}}.
    (\bibinfo{year}{2017}).
\newblock


\bibitem[\protect\citeauthoryear{Owen}{Owen}{2015}]%
        {Owen:2015}
\bibfield{author}{\bibinfo{person}{Taylor Owen}.}
  \bibinfo{year}{2015}\natexlab{}.
\newblock \bibinfo{booktitle}{{\em Disruptive Power: The Crisis of the State in
  the Digital Age}}.
\newblock \bibinfo{publisher}{Oxford Studies in Digital Politics, New York}.
\newblock


\bibitem[\protect\citeauthoryear{Pathak and Nanded}{Pathak and Nanded}{2016}]%
        {Pathak:2016a}
\bibfield{author}{\bibinfo{person}{P.B. Pathak} {and}
  \bibinfo{person}{Yeshwant~Mahavidyalaya Nanded}.}
  \bibinfo{year}{2016}\natexlab{}.
\newblock \showarticletitle{A dangerous Trend of Cybercrime: Ransomware Growing
  Challenge}.
\newblock \bibinfo{journal}{{\em International Journal of Advanced Research in
  Computer Engineering \& Technology (IJARCET) Volume\/}}  \bibinfo{volume}{5}
  (\bibinfo{year}{2016}).
\newblock


\bibitem[\protect\citeauthoryear{Reid and Harrigan}{Reid and Harrigan}{2013}]%
        {Reid:2013a}
\bibfield{author}{\bibinfo{person}{Fergal Reid} {and} \bibinfo{person}{Martin
  Harrigan}.} \bibinfo{year}{2013}\natexlab{}.
\newblock \showarticletitle{An analysis of Anonymity in the Bitcoin system}.
\newblock In \bibinfo{booktitle}{{\em Security and privacy in social
  networks}}. \bibinfo{publisher}{Springer}, \bibinfo{pages}{197--223}.
\newblock


\bibitem[\protect\citeauthoryear{Scaife, Carter, Traynor, and Butler}{Scaife
  et~al\mbox{.}}{2016}]%
        {Scaife:2016a}
\bibfield{author}{\bibinfo{person}{Nolen Scaife}, \bibinfo{person}{Henry
  Carter}, \bibinfo{person}{Patrick Traynor}, {and} \bibinfo{person}{Kevin~RB
  Butler}.} \bibinfo{year}{2016}\natexlab{}.
\newblock \showarticletitle{Cryptolock (and drop it): Stopping Ransomware
  Attacks on User Data}. In \bibinfo{booktitle}{{\em Distributed Computing
  Systems (ICDCS), 2016 IEEE 36th International Conference on}}. IEEE,
  \bibinfo{pages}{303--312}.
\newblock


\bibitem[\protect\citeauthoryear{Scott and Drew}{Scott and Drew}{2016}]%
        {Scott:2016:Online}
\bibfield{author}{\bibinfo{person}{James Scott} {and} \bibinfo{person}{Spaniel
  Drew}.} \bibinfo{year}{2016}\natexlab{}.
\newblock \bibinfo{booktitle}{{\em The ICIT Ransomware Report: 2016 will be the
  Year Ransomware holds America Hostage}}.
\newblock \bibinfo{type}{{T}echnical {R}eport}. \bibinfo{institution}{Institute
  for Critical Infrastructure Technology}.
\newblock
\showURL{%
\url{http://icitech.org/wp-content/uploads/2016/03/ICIT-Brief-The-Ransomware-Report2.pdf}}


\bibitem[\protect\citeauthoryear{Song, Kim, and Lee}{Song
  et~al\mbox{.}}{2016}]%
        {Song:2016a}
\bibfield{author}{\bibinfo{person}{Sanggeun Song}, \bibinfo{person}{Bongjoon
  Kim}, {and} \bibinfo{person}{Sangjun Lee}.} \bibinfo{year}{2016}\natexlab{}.
\newblock \showarticletitle{The Effective Ransomware Prevention Technique Using
  Process Monitoring on Android Platform}.
\newblock \bibinfo{journal}{{\em Mobile Information Systems\/}}
  \bibinfo{volume}{2016} (\bibinfo{year}{2016}).
\newblock


\bibitem[\protect\citeauthoryear{Spagnuolo, Maggi, and Zanero}{Spagnuolo
  et~al\mbox{.}}{2014}]%
        {Spagnuolo:2014b}
\bibfield{author}{\bibinfo{person}{Michele Spagnuolo},
  \bibinfo{person}{Federico Maggi}, {and} \bibinfo{person}{Stefano Zanero}.}
  \bibinfo{year}{2014}\natexlab{}.
\newblock \showarticletitle{Bitiodine: Extracting Intelligence from the Bitcoin
  Network}. In \bibinfo{booktitle}{{\em International Conference on Financial
  Cryptography and Data Security}}. Springer, \bibinfo{pages}{457--468}.
\newblock


\bibitem[\protect\citeauthoryear{Wall}{Wall}{2007}]%
        {Wall:2007a}
\bibfield{author}{\bibinfo{person}{David Wall}.}
  \bibinfo{year}{2007}\natexlab{}.
\newblock \bibinfo{booktitle}{{\em Cybercrime: The transformation of crime in
  the information age}}. Vol.~\bibinfo{volume}{4}.
\newblock \bibinfo{publisher}{Polity}.
\newblock


\bibitem[\protect\citeauthoryear{Whitepaper}{Whitepaper}{2016a}]%
        {Kaspersky:2016:Online}
\bibfield{author}{\bibinfo{person}{Kaspersky Whitepaper}.}
  \bibinfo{year}{2016}\natexlab{a}.
\newblock \bibinfo{booktitle}{{\em Security Bulletin: Overall Statistics for
  2016}}.
\newblock \bibinfo{type}{{T}echnical {R}eport}.
  \bibinfo{institution}{Kaspersky}.
\newblock
\showURL{%
\url{https://kasperskycontenthub.com/securelist/files/2016/12/Kaspersky_Security_Bulletin_2016_Statistics_ENG.pdf}}


\bibitem[\protect\citeauthoryear{Whitepaper}{Whitepaper}{2016b}]%
        {RSA:2016:Online}
\bibfield{author}{\bibinfo{person}{RSA Whitepaper}.}
  \bibinfo{year}{2016}\natexlab{b}.
\newblock \bibinfo{booktitle}{{\em 2016: Current State of Cybercrime}}.
\newblock \bibinfo{type}{{T}echnical {R}eport}. \bibinfo{institution}{RSA}.
\newblock
\showURL{%
\url{https://www.rsa.com/content/dam/rsa/PDF/2016/05/2016-current-state-of-cybercrime.pdf}}


\end{thebibliography}

\end{document}